\documentclass[12pt]{article}
\usepackage{amsfonts}
\usepackage{amsmath}
\usepackage{latexsym}
\usepackage{epsfig,epstopdf}
\def\beq{\begin{equation}}
\def\eeq{\end{equation}}
\def\bea{\begin{eqnarray}}
\def\eea{\end{eqnarray}}
\begin{document}

\begin{center}
{\Large \bf Can the 750 GeV diphoton LHC excess be due to a radion-dominated state?} \\

\vspace{4mm}

Edward E.~Boos, Viacheslav E.~Bunichev,  Igor P.~Volobuev \\
\vspace{4mm}
 Skobeltsyn Institute of Nuclear Physics,
Moscow State University
\\ 119991 Moscow, Russia \\

\end{center}

\begin{abstract}
We discuss the possibility of  interpreting the 125 GeV scalar
boson and the 750 GeV diphoton excess recently  reported by the
ATLAS and CMS experiments  as a Higgs-dominated and  a
radion-dominated states respectively in a stabilized brane-world
model. It is shown that in the simplest variant of the model,
where only the  gravitational degrees of freedom  propagate in the
bulk, the production cross section of the radion-dominated state
with mass 750 GeV  turns out to be too small in the allowed region
of the model parameter space for explaining the nature of the
excess in this approach.

\end{abstract}

\section{Introduction}
The 750 GeV diphoton  excess that has been recently presented by
the LHC experiments \cite{ATLAS, CMS} can be the first direct
evidence of physics beyond the Standard Model (SM), if it is
confirmed by further searches at larger statistics. The most
plausible explanation of this excess seems to be the production of
a new heavy scalar state \footnote{A vector state cannot  decay to
two photons because of the Landau-Yang theorem. Spin two states
can. However, if they are somehow related to the graviton
excitations similar excesses should appear in di-lepton and di-jet
decay modes, which are not seen in the LHC experiments. }. There
have been already numerous attempts either to explain the excess
in a model independent way (see, e.g., \cite{generic-resonance}),
or to associate it with the  scalar states predicted by concrete
SM extensions (see, e.g.,
\cite{resonance-in-models,Gupta:2015zzs}), in particular, to
interpret it as a mixed Higgs-radion state
\cite{Cox:2015ckc,Gunion,Megias:2015ory} in the Randall-Sundrum
model \cite{RS}.

In the present paper we consider the case, where the two scalar
states are interpreted as  mixed Higgs-radion states in  an
extension of the SM based on the Randall-Sundrum model with two
branes stabilized by a bulk scalar field
\cite{wise,wolfe,Boos:2004uc,Boos:2005dc}, which is necessary for
the model to be phenomenologically acceptable. A characteristic
feature of this extension is the presence of a massive scalar
radion field together with its Kaluza-Klein tower. These fields
have the same quantum numbers as the neutral Higgs field and can
mix with the Higgs field, if they are coupled.  It is worth noting
that in the model under consideration all the SM fields are
located on the TeV brane, which does not lead to any contradiction
with the Electroweak Precision Data
\cite{Csaki:2000zn,Csaki:2002gy,Burdman:2002se}.

Usually, the Higgs-radion coupling in  the Randall-Sundrum model
is obtained  by introducing a Higgs-curvature term on the brane
\cite{Giudice:2000av}. In paper \cite{Boos:2015xma} a model was
discussed, where  a Higgs-radion coupling  arises due to a
mechanism of spontaneous symmetry breaking on the brane involving
the stabilizing scalar field. This approach takes into account the
influence of the  KK tower of higher scalar excitations on the
parameters of the Higgs-radion mixing, which turns out to be of
importance. It also has the advantage that, unlike the approach
based on the Higgs-curvature term, it modifies only the scalar
sector of the model and leaves intact the coupling constants and
the masses  of the graviton KK excitations.  In this case the
latter are  of the order of the fundamental five-dimensional
energy scale. In the present paper we take this energy scale  to
be 2 TeV or larger. Therefore, the  masses  of the graviton KK
excitations should also be larger than 2 TeV and  the absence of
any excess in the di-lepton mode in the region of 750 GeV is quite
consistent with the property of the stabilized brane world  models
to have the KK graviton states that are much heavier than the
radion.

Here we will use the effective interaction Lagrangian obtained in
paper \cite{Boos:2015xma} to describe the couplings of the mixed
Higgs-radion states to the SM fields and to analyze the
interpretation  of the excess as a  750 GeV mixed Higgs-radion
scalar state.

\section{Higgs-radion interaction and the effective Lagrangian}
First let us  briefly recall the main features of the model
(details are given in \cite{Boos:2015xma}).  It  is a variant of
the stabilized RS model with all the SM fields located on the TeV
brane \cite{Rubakov:2001kp,Giudice:2004mg,Boos:2004uc}  and the
stabilizing Goldberger-Wise scalar field propagating in the bulk
along with the gravitational field. The mixing of the Higgs and
gravitational scalar fields arises in the model due to an
interaction Lagrangian leading  to a relation between the Higgs
vev and the value of the stabilizing field on the UV brane that
correspond to the minimum of the potential.

The model  in five-dimensional
space-time $E=M_4\times S^{1}/Z_{2}$  with coordinates $\{ x^M\}
\equiv \{x^{\mu},y\}$, $M = 0,1,2,3,4, \, \mu=0,1,2,3 $, the
coordinate $x^4 \equiv y, \quad -L\leq y \leq L$ parameterizing
the fifth dimension is  defined by the   action
\begin{equation}\label{actionDW}
S = S_g + S_{\phi+SM},
\end{equation}
where $S_g$ and $S_{\phi+SM}$ are given by
\begin{eqnarray}\label{actionsDW}
S_g&=& - 2 M^3\int d^{4}x \int_{-L}^L dy  R\sqrt{g},\\
\label{actionsSM} S_{\phi+SM} &=& \int d^{4}x \int_{-L}^L dy
\left(\frac{1}{2}
g^{MN}\partial_M\phi\partial_N\phi -V (\phi)\right)\sqrt{g} -\\
\nonumber & -&\int_{y=0} \sqrt{-\tilde g}V_1(\phi) d^{4}x
+\int_{y=L}\sqrt{-\tilde g}(-V_2(\phi)+ L_{SM-HP} + L_{int}(\phi,
H))    d^{4}x .
\end{eqnarray}
The signature of the metric  $g_{MN}$ is chosen to be
$(+,-,-,-,-)$, $M$ is the fundamental five-dimensional energy
scale, $V(\phi)$ is a bulk stubilizing scalar field potential and
$V_{1,2}(\phi)$ are brane scalar field potentials,
$\tilde{g}=det\tilde g_{\mu\nu}$, and $\tilde g_{\mu\nu}$  denotes
the metric induced on the branes.  The space of extra dimension is
the orbifold  $S^{1}/Z_{2}$, which is realized as the circle of
circumference $2L$ with the points $y$ and $-y$ identified.
Correspondingly, the metric $g_{MN}$  and the scalar field $\phi$
satisfy the standard orbifold symmetry conditions and the branes
are located at the fixed points of the orbifold, $y=0$ and $y=L$.
The SM fields are assumed to be localized on the brane at $y = L$,
and the Lagrangian $L_{SM-HP}$ is the SM Lagrangian without the
Higgs potential. The key point of the  approach under
consideration is the replacement of the Higgs potential  by the
interaction Lagrangian
\begin{equation}\label{L_int} L_{int}(\phi, H) = -\lambda(|H|^2 -\frac{\xi}{M} \phi^2)^2
\end{equation}
of the Higgs  and  Goldberger-Wise fields,   $\xi$ being a
positive dimensionless parameter.

The background solutions  for the  metric and the scalar field,
which preserve the Poincar\'e invariance in any four-dimensional
subspace $y=const$, look like
\begin{eqnarray}\label{metricDW}
ds^2&=&  e^{-2A(y)}\eta_{\mu\nu}  {dx^\mu  dx^\nu} -  dy^2 \equiv
\gamma_{MN}(y)dx^M dx^N, \\
\phi(x,  y) &=& \phi(y),
\end{eqnarray}
$\eta_{\mu\nu}$ denoting the flat Minkowski metric, whereas the
background (vacuum) solution for the Higgs field is standard
\begin{equation}\label{Higgs_vac}
 H_{vac}= \begin{pmatrix}
0\\
\frac{v}{\sqrt{2}}
\end{pmatrix},
\end{equation}
all the other SM fields  being equal to zero.

 If one substitutes this ansatz into the equations corresponding
to action (\ref{actionDW}), one gets  a system of differential
equations for the functions $A(y)$ and $\phi(y)$, the brane scalar
field potentials $V_{1,2}(\phi)$ and  interaction Lagrangian
(\ref{L_int}) defining the boundary conditions for these equations
on the branes. The potentials $V_{1,2}(\phi)$ are chosen so that
they fix the values of the stabilizing field $\phi$ on the branes
and stabilize the interbrane distance \cite{wise,wolfe}.
Interaction Lagrangian (\ref{L_int}) does not affect this
stabilization mechanism, if the relation
\begin{equation}\label{relation}
\phi(L)^2 = \frac{M v^2}{2\xi}
\end{equation}
between the values of the Higgs and stabilizing scalar field on
the brane at $y = L $ is valid, which defines the vacuum value of
the Higgs field. This means that in such a scenario the Higgs
field vacuum expectation value, being proportional to the value of
the stabilizing scalar field on the TeV brane, arises dynamically
as a result of the gravitational bulk stabilization.

Now the linearized theory is obtained by expanding the metric,
the scalar and the Higgs field in the unitary gauge around the
background solution as
\begin{eqnarray}\label{metricparDW}
g_{MN}(x,y)&=& \gamma_{MN}(y) + \frac{1}{\sqrt{2M^3}} h_{MN}(x,y),
\\ \label{metricparDW1}
\phi(x,y) &=& \phi(y) + \frac{1}{\sqrt{2M^3}} f(x,y),\\
 H(x)&=& \begin{pmatrix}
0\\
\frac{v+ \sigma(x)}{\sqrt{2}}
\end{pmatrix}.
\end{eqnarray}
After  substituting this representation into action
(\ref{actionDW}) and keeping the terms of the second order  in
$h_{MN}$, $f$ and $\sigma$ one gets the Lagrangian of this action
which is the standard free Lagrangian of the SM (i.e. the masses
of all the SM fields are expressed in the same way as usually in
terms of the vacuum value of the Higgs field and the coupling
constants) together with  the standard second variation Lagrangian
of the stabilized RS model \cite{Boos:2005dc} supplemented by an
interaction term of  the scalar  fields $f$ and $\sigma$  on the
brane coming from interaction Lagrangian (\ref{L_int})
\cite{Boos:2015xma}.

 Besides the fields $f$ and $\sigma$, there two more scalar
fields in the linearized theory, --  the fields $h_{44}(x,y)$ and
$\gamma^{\mu\nu} h_{\mu\nu}(x,y)$. However, the fields  $f(x,y)$,
$h_{44}(x,y)$ and $\gamma^{\mu\nu} h_{\mu\nu}(x,y)$ are not
independent: they are connected  by the equations of motion of the
linearized theory and a gauge condition
\cite{Boos:2005dc,Csaki:2000zn}. For this reason we can use any
one of them to describe the scalar states.

The field $h_{44}(x,y)$ can be expanded in KK modes, the lowest
mode $\phi_1(x)$ is called the radion field and the modes
$\phi_n(x), \quad n > 1$ belong to its KK tower. This expansion
induces the corresponding expansion of the bulk scalar field
$f(x,y)$. Substituting the latter expansion into the second
variation Lagrangian and integrating over the extra dimension
coordinate, one gets an effective four-dimensional Lagrangian. In
case the Higgs and the radion masses are  much lower  than the
masses of the radion excitations on can pass to a low energy
approximation in the four-dimensional Lagrangian  by integrating
out the radion excitation fields. This gives an effective
Lagrangian for the interactions of the Higgs and radion fields
with the SM fields. However, due to the Higgs-radion mixing terms
the fields $\sigma(x)$ and $\phi_1(x)$ are not mass eigenstates.

The physical mass eigenstate fields $h(x), r(x)$ are, as usually,
obtained by a rotation diagonalizing the mass matrix
\begin{eqnarray}\label{phys_states}
h(x) &=& \cos \theta \, \sigma(x) + \sin \theta \,\phi_1(x)\\
\nonumber r(x) &=& -  \sin \theta \,\sigma(x) + \cos \theta
\,\phi_1(x).
\end{eqnarray}
The field $h(x)$ is  the Higgs-dominated field and the field
$r(x)$ is the radion-dominated field. Finally, one gets the
effective interaction Lagrangian  of the physical scalar fields
$h(x)$ and $r(x)$ with the Standard Model fields in the following
form \cite{Boos:2015xma}:

\begin{eqnarray}\label{h-r-Lagrangian}
L_{h-r} &=& \frac{1}{2} \partial_\mu h(x) \partial^ \mu h(x) -
\frac{1}{2} m^2_h h^2(x) + \frac{1}{2}
\partial_\mu r(x) \partial^\mu r(x) -
\frac{1}{2} \mu_r^2 r^2(x) \\ \nonumber &-&  \frac{(c \cos \theta
+ \sin \theta)}{\Lambda_r} h(x) (T_\mu^\mu + \Delta T_{\mu}^{\mu})
+ \frac{(c\sin  \theta -
\cos\theta)}{\Lambda_r} r(x)(T_\mu^\mu + \Delta T_{\mu}^{\mu}) -\\
\nonumber&-& \sum_{f} \frac{m_f}{v} \bar \psi_f \psi_f (\cos
\theta \, h(x) - \sin \theta \,r(x))+ \frac{2 M^2_W}{v} (W_\mu^-
W^{\mu +})(\cos \theta \,h(x) - \sin \theta \,r(x)) + \\
\nonumber&+& \frac{M^2_Z}{v} (Z_\mu Z^\mu)(\cos \theta \, h(x) -
\sin \theta \, r(x)) + \frac{M^2_W}{v^2} (W_\mu^- W^{\mu
+})(\cos \theta \, h(x) - \sin \theta \, r(x))^2 + \\
\nonumber&+& \frac{M^2_Z}{2 v^2} (Z_\mu Z^\mu)(\cos \theta \, h(x)
- \sin \theta \, r(x))^2.
\end{eqnarray}

Here  $m_h^2$  and $m_r^2$ are the masses of the fields  $h(x)$
and $r(x)$,  $\Lambda_r$ is the (inverse) coupling constant of the
radion to the trace of the SM energy-momentum tensor, $\Delta
T_{\mu}^{\mu}$ is the conformal anomaly of massless vector fields
explicitly given by
\begin{equation}
\Delta T_{\mu}^{\mu} =
\frac{\beta(g_s)}{2g_s}G^{ab}_{\rho\sigma}G_{ab}^{\rho\sigma} +
\frac{\beta(e)}{2e}F_{\rho\sigma}F^{\rho\sigma}
 \label{Tr1}
\end{equation}
with  $\beta$ being the well-known QCD and QED $\beta$-functions.

The parameter $c$  accommodates the contributions of the
integrated out heavy scalar modes and is expressed  in terms
of the physical masses and the mixing angle as follows:
\begin{equation}\label{restriction}
 c =  \frac{(m_r^2 - m_h^2)\sin 2
 \theta}{m_r^2 \cos^2 \theta + m_h^2 \sin^2 \theta } \left(\sum_{n=2}^{\infty}
 \alpha_n^2\right).
\end{equation}
It also depends on the sum of the coefficients $\alpha_n^2$, where
$\alpha_n$ is  the ratio of the wave functions in the extra
dimension of the modes $\phi_n$ and $\phi_1$ taken at $y = L$ .
These ratios are, of course, model dependent and  should fall off
with $n$ in order for the sum to be convergent. In certain models
several first ratios may be of the order of unity
\cite{Boos:2005dc}. Thus, one can conservatively estimate this sum
to be also of the order of unity.

The effective four-dimensional  interaction Lagrangian
(\ref{h-r-Lagrangian}) expressed in terms of the physical
Higgs-dominated  $h(x)$ and radion-dominated $r(x)$ fields
involves only five extra parameters in addition to those of the
SM: the masses of the Higgs-dominated and radion-dominated fields
$m_h$ and $m_r$, the mixing angle $\theta$, the (inverse) coupling
constant of the radion to the trace of the energy-momentum tensor
of the SM fields $\Lambda_r$ and the parameter $c$.

Let us consider several interesting subspaces of the parameter
space of the effective theory.  If we put  the parameters  $c$ and
$\theta$ equal to zero, i.e. consider the case of the zero mixing,
Lagrangian (\ref{h-r-Lagrangian}) becomes just the SM Higgs
Lagrangian plus the usual Lagrangian of the radion interaction
with  the trace of the  SM energy-momentum tensor.   If we put $1/
\Lambda_r = 0$, we get the effective Lagrangian of the real Higgs
singlet extension of the Standard Model \cite{Robens:2015gla,
Godunov:2015nea}. If we formally put  the parameter $c$ equal to
zero while keeping the mixing angle $\theta$ non-zero, we obtain
an effective interaction Lagrangian which is very similar to the
one of the unstabilized RS model with the Higgs-curvature term on
the brane \cite{Giudice:2000av}. In particular, the couplings of
the Higgs-dominated and the radion-dominated states to the
conformal anomaly of massless vector fields, which turn to be very
important for their production and decay to two photons, are the
same. However, one should keep in mind that the observable
parameters of the effective Lagrangian in different cases depend
differently on the fundamental parameters of the models.

Thus, Lagrangian (\ref{h-r-Lagrangian}) is   a very  general
effective Lagrangian of the interaction of two mixed scalar states
with the fermion and vector fields of the Standard Model extended
by a singlet scalar. The extra parameters of this Lagrangian are
natural for considering the phenomenology of the mixed scalar
states and allow one to compare easily the predictions of
different models. In fact, it is  unimportant, how they depend on
the fundamental parameters of a particular model as long as the
latter belong to the phenomenologically acceptable parameter
subspace.

In the general case of  a non-zero mixing, when all the parameters
and $\theta$,  $1/ \Lambda_r$, and  $c$   are not equal to zero,
the additional terms in the Lagrangian, coming from the integrated
out heavy modes and containing the parameter $c$ that depends on
the scalar state masses and the mixing angle,  may lead to certain
changes in the collider phenomenology of the Higgs-dominated and
radion-dominated states as was demonstrated in
\cite{Boos:2015xma}, where also the Feynman rules for the model
are explicitly given.

\section{Phenomenology of the 750 GeV radion-dominated state}

In order to understand whether or not the 750 GeV observed excess
may be interpreted as a radion-dominated state in the  above
described stabilized brane world model let us consider the main
decay and production properties of such a state as follow from
effective Lagrangian (\ref{h-r-Lagrangian}). For the computations
a special version of the CompHEP code \cite{Boos:2004kh,
Boos:2009un} was used in the same manner as was done in
\cite{Boos:2015xma}. The version includes a special routine for
$\chi^2$-analysis of signal strengths and  implements the Feynman
rules corresponding to effective Lagrangian
(\ref{h-r-Lagrangian}).

\begin{figure*}[!h!]
\begin{center}

\begin{minipage}[t]{.45\linewidth}
\centering
\includegraphics[width=80mm,height=70mm]{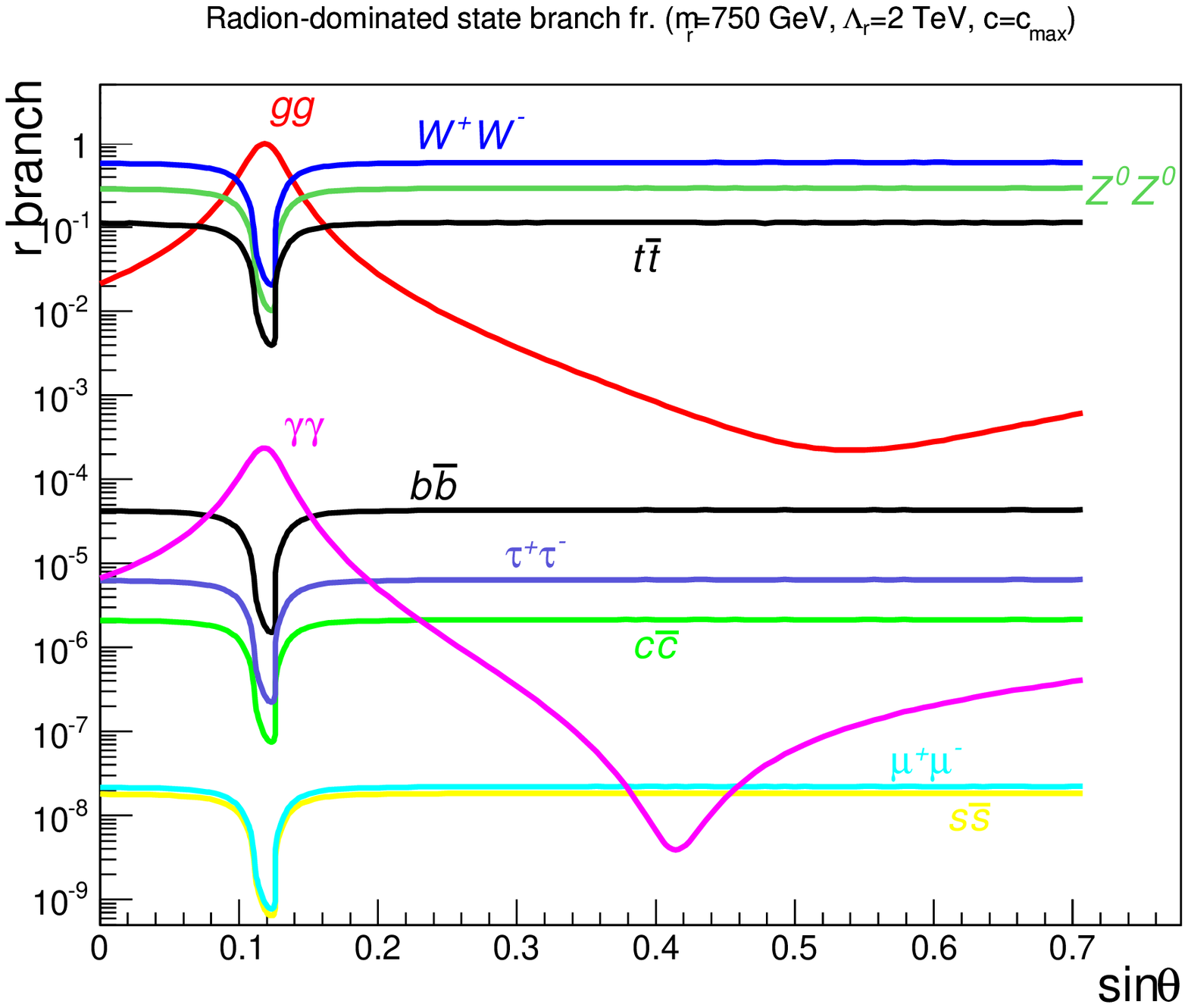}
\caption{\label{pic1} \footnotesize The decay branching ratios for
the radion-dominated state with  mass 750 GeV  as functions of the
mixing angle parameter $\sin \theta$. }
\end{minipage}
\hfill
\begin{minipage}[t]{.45\linewidth}
\centering
\includegraphics[width=80mm,height=70mm]{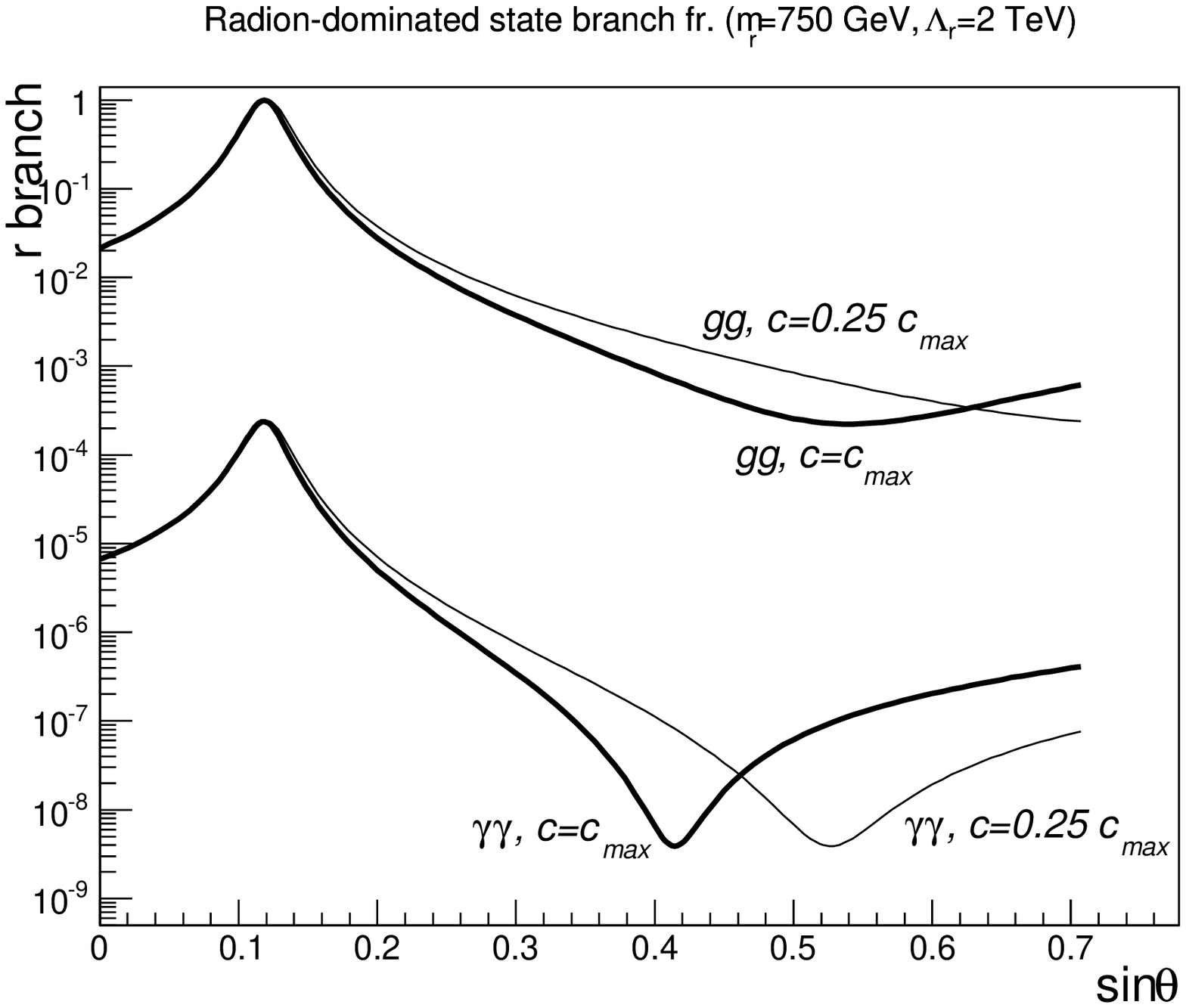}
\caption{ \label{pic2} \footnotesize The gluon-gluon and
photon-photon branchings as functions of $\sin \theta$ for $c =
c_{max}$ and  $c = 0.25 c_{max}$.}
\end{minipage}

\end{center}
\end{figure*}
 The decay branching ratios for
the radion-dominated state with mass 750 GeV are shown in
Fig.~\ref{pic1} as functions of the mixing angle parameter $\sin
\theta$. One can see that practically for all the values of the
mixing parameter $\sin \theta$ the branching ratios are
distributed between the modes close to those for the SM Higgs
boson, if it would have had  mass 750 GeV.  The NLO corrections
are included following the HDECAY code \cite{hdecay}. The
dominating decay modes are  the decays to heavy SM particles
$W^+W^-, ZZ$ boson pairs and the top quark pair. However, for some
rather small values of the  parameter $\sin \theta$  close to
approximately $v/\Lambda_r$ all the branching ratios are
significantly decreased, and the dominating decay mode becomes the
mode to two gluons. Also in this region of the parameter space the
branching to two photons is significantly increased. Such a
property could be easily understood from the structure of the
interaction vertices of the radion-dominated state and the SM
fermions and gauge bosons (see the Feynman rules in
\cite{Boos:2015xma}). Indeed all the vertices for the fermions and
massive gauge bosons contain the factor $\frac{\cos\theta-c\cdot
\sin\theta}{\Lambda_r}-\frac{\sin\theta}{v}$, which becomes very
small for $\sin\theta$ close to $v/\Lambda_r$. This occurs due to
the cancellation of the  contributions to the vertices coming from
the SM type part of the interactions and the part coming from the
trace of the energy-momentum tensor. In contrast, the interaction
vertices of the radion-dominated state and the  massless gluons
and photons have anomaly enhanced contributions and the  mentioned
cancellation does not take place for small values of the parameter
$\sin \theta$ close to  $v/\Lambda_r$. The corresponding
cancellation occurs for the gluon-gluon and photon-photon vertices
at much larger values of the  parameter $\sin \theta$, where the
branching ratios have minima as can be seen in  Fig.~\ref{pic1}.
One should mention that the position of the maximum value for the
gluon and photon decay modes goes to smaller and smaller values of
$\sin \theta$ with the  increase of the scale parameter
$\Lambda_r$. The position of the maximum as well as the form of
the curves close to the maximum practically does not depend on the
value of the parameter $c$, which accumulates the contributions of
the higher KK scalar modes. The dependence on the parameter $c$
for two different values of this parameter is demonstrated in
Fig.~\ref{pic2} showing the gluon-gluon and photon-photon
branchings as functions of $\sin \theta$.

\begin{figure*}[!h!]
\begin{center}

\begin{minipage}[t]{.45\linewidth}
\centering
\includegraphics[width=80mm,height=70mm]{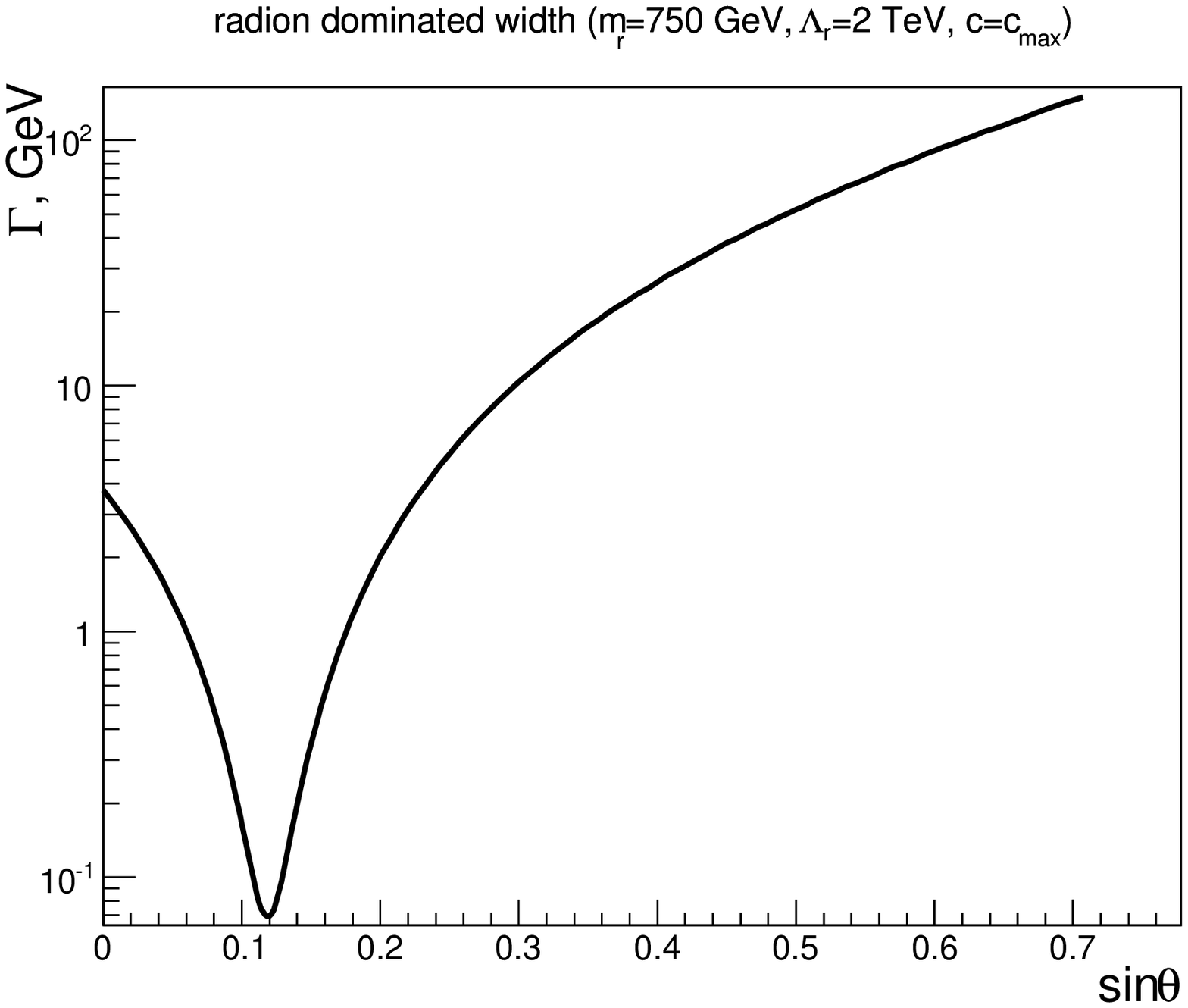}
\caption{\label{pic3} \footnotesize The total width of the
radion-dominated state with  mass 750 GeV  as a function of the
mixing angle parameter $\sin \theta$. }
\end{minipage}
\hfill
\begin{minipage}[t]{.45\linewidth}
\centering
\includegraphics[width=80mm,height=70mm]{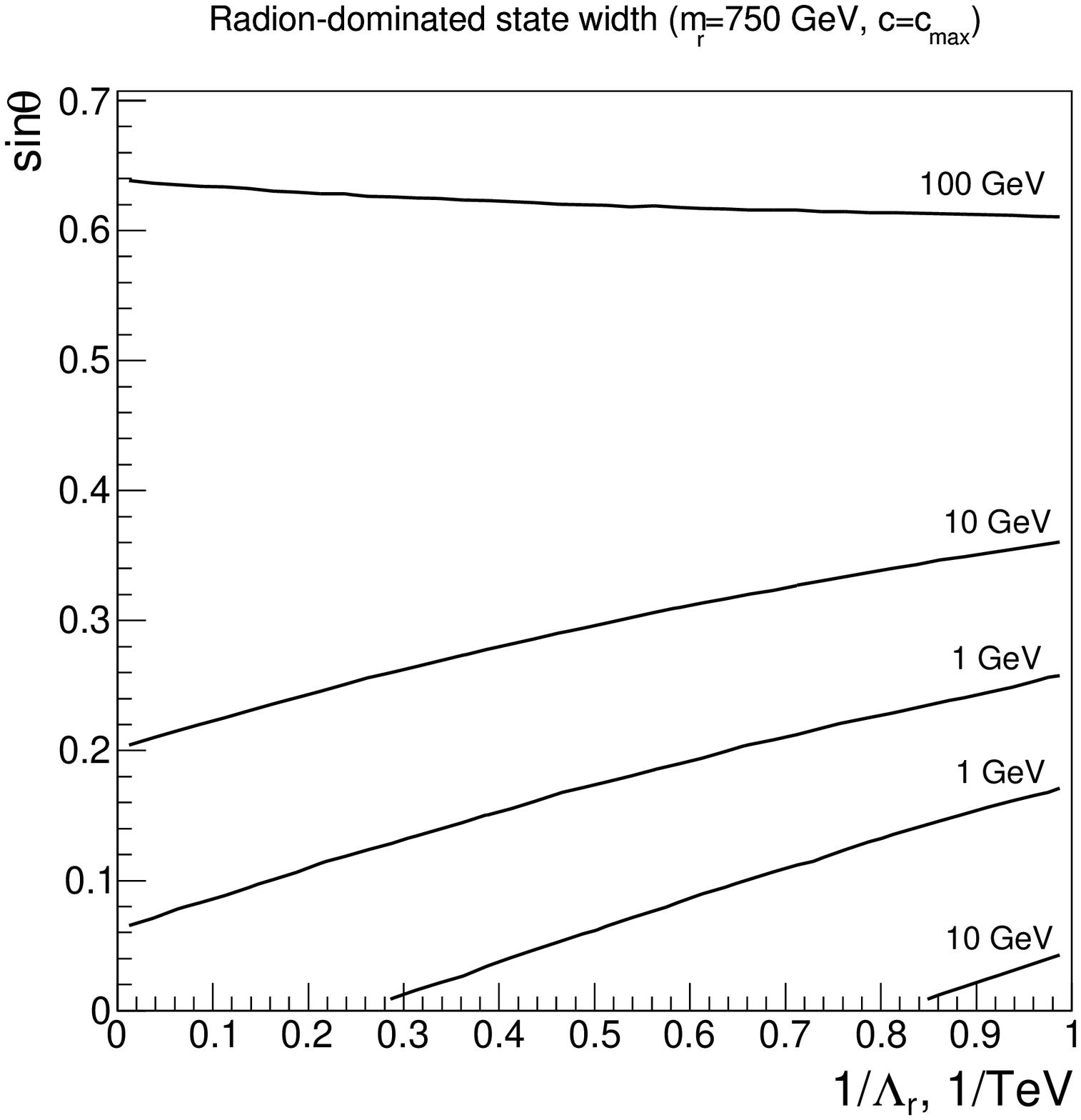}
\caption{ \label{pic4} \footnotesize Equal width contours  for the
radion-dominated state with  mass 750 GeV  as functions of the
parameters $\sin \theta$ and $1/\Lambda_r$.}
\end{minipage}

\end{center}
\end{figure*}
Fig.~\ref{pic3} illustrates the behavior of the total width of the
radion-dominated state. The deep minimum  corresponds exactly to
the discussed region, where all the leading decay modes
drastically go down. For a particular choice of the parameters
$\Lambda_r = 2$ TeV and  $c = c_{max}$ the width in the minimum
becomes  as small as $6 \cdot 10^{-2}$ GeV. However, for the
values of the parameter $\sin \theta$ away from the minimum region
the width can be significantly larger reaching a few tens or even
a hundred GeV. In the two-dimensional plot (see Fig.~\ref{pic4})
the lines of equal width values are shown depending on the
parameters $\sin \theta$ and $\Lambda_r$. The plot demonstrates
that the radion-dominated state can be rather wide. The width of
the 750 GeV excess observed at the LHC has a rather large value of
the order of 45 GeV \cite{ATLAS, CMS}.

\begin{figure*}[!h!]
\begin{center}

\begin{minipage}[t]{.45\linewidth}
\centering
\includegraphics[width=80mm,height=70mm]{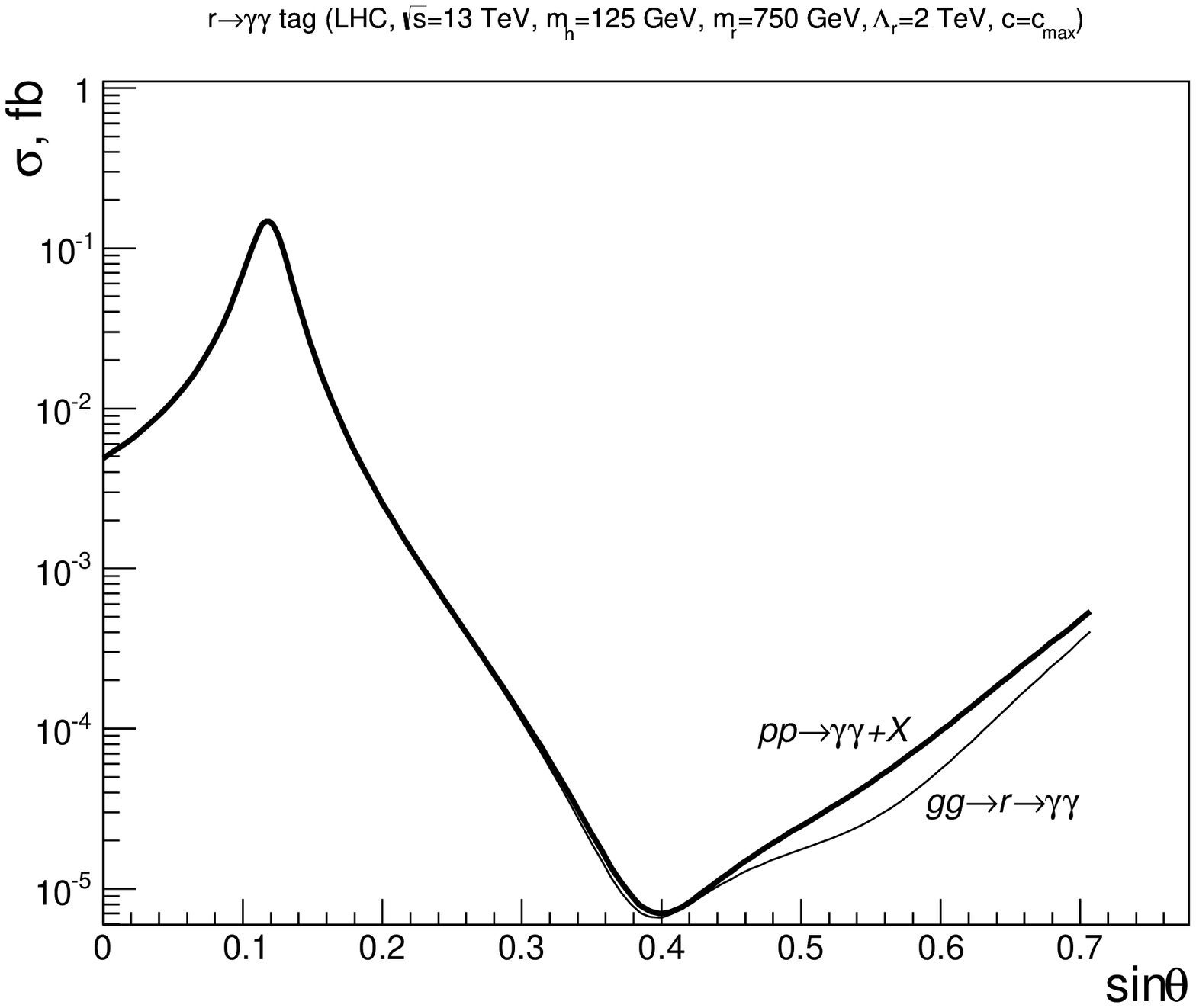}
\caption{\label{pic5} \footnotesize The production cross section
of the radion-dominated state with  mass 750 GeV  as a function of
the mixing angle parameter $\sin \theta$ including the
contributions of all the production modes (thick curve) and only
the leading contribution of the gluon-gluon fusion mode (thin
curve) }
\end{minipage}
\hfill
\begin{minipage}[t]{.45\linewidth}
\centering
\includegraphics[width=80mm,height=70mm]{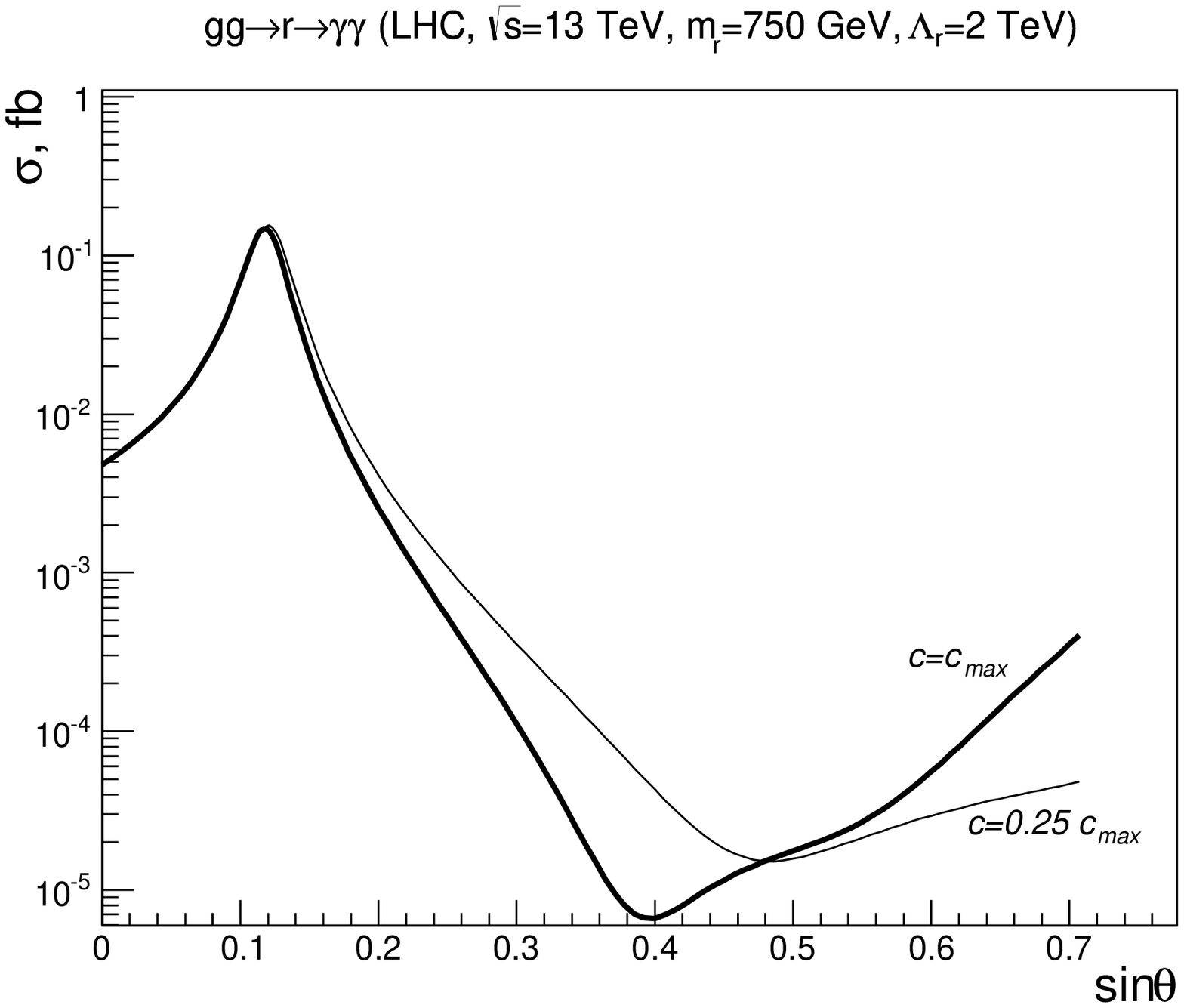}
\caption[]{\label{pic6} \footnotesize The production cross
section of the radion-dominated state in the main gluon-gluon fusion mode with  mass 750 GeV  as a
function of the mixing angle parameter $\sin \theta$ for the
parameter $c = c_{max}$ (thick curve) and $c = 0.25 c_{max}$ (thin curve).}
\end{minipage}

\end{center}
\end{figure*}

However, a crucial point for the possible interpretation is the
production cross section, which should be in the range from a few
to 10 fb. The production cross section for the radion-dominated
state is shown in  Fig.~\ref{pic5}. We have included the
contributions of all the production channels for the
radion-dominated state (ggF, VBF, rV, rtt) with the decay to two
photons. We have included the NNLO K-factors taken from the Higgs
cross section working group web page
(\cite{Dittmaier:2011ti,Heinemeyer:2013tqa}). One can see that, as
expected, the gluon-gluon fusion dominates the production cross
section in the most interesting region of the parameter space,
where the cross section has a maximum. The maximum occurs for the
same values of parameters, for which the  corresponding gg and
$\gamma\gamma$ decay branching ratios have the maximum. As for the
branching ratios in the range close to the maximum, the cross
section depends very weakly on the the parameter $c$. Such a
dependence is significant for rather large mixing angles, where
the cross section becomes much smaller (see Fig.~\ref{pic6}).

\begin{figure*}[!h!]
\begin{center}

\begin{minipage}[t]{.45\linewidth}
\centering
\includegraphics[width=80mm,height=70mm]{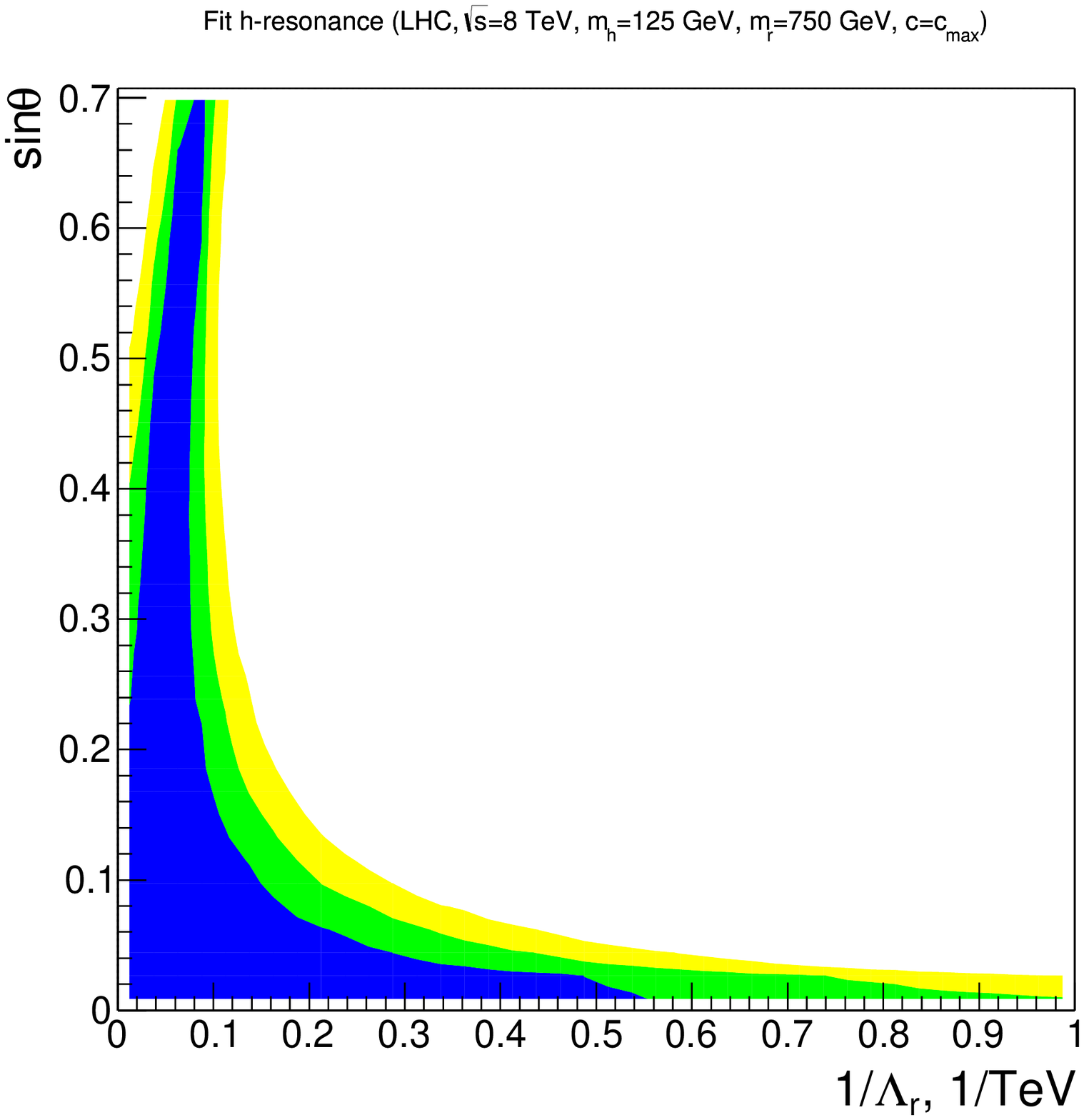}
\caption{\label{pic7} \footnotesize  Exclusion contours for the
global $\chi^2$ fit in the $(\sin\theta, 1/\Lambda_r)$ plane for
the LHC at $\sqrt{s}=$7 and 8 TeV and  $m_h$=125 GeV, $m_r$=750
GeV, $c=c_{max}$. The dark, medium and light shaded areas
correspond to CL of the fit 65\%, 90\% and 99\% respectively. }
\end{minipage}
\hfill
\begin{minipage}[t]{.45\linewidth}
\centering
\includegraphics[width=80mm,height=70mm]{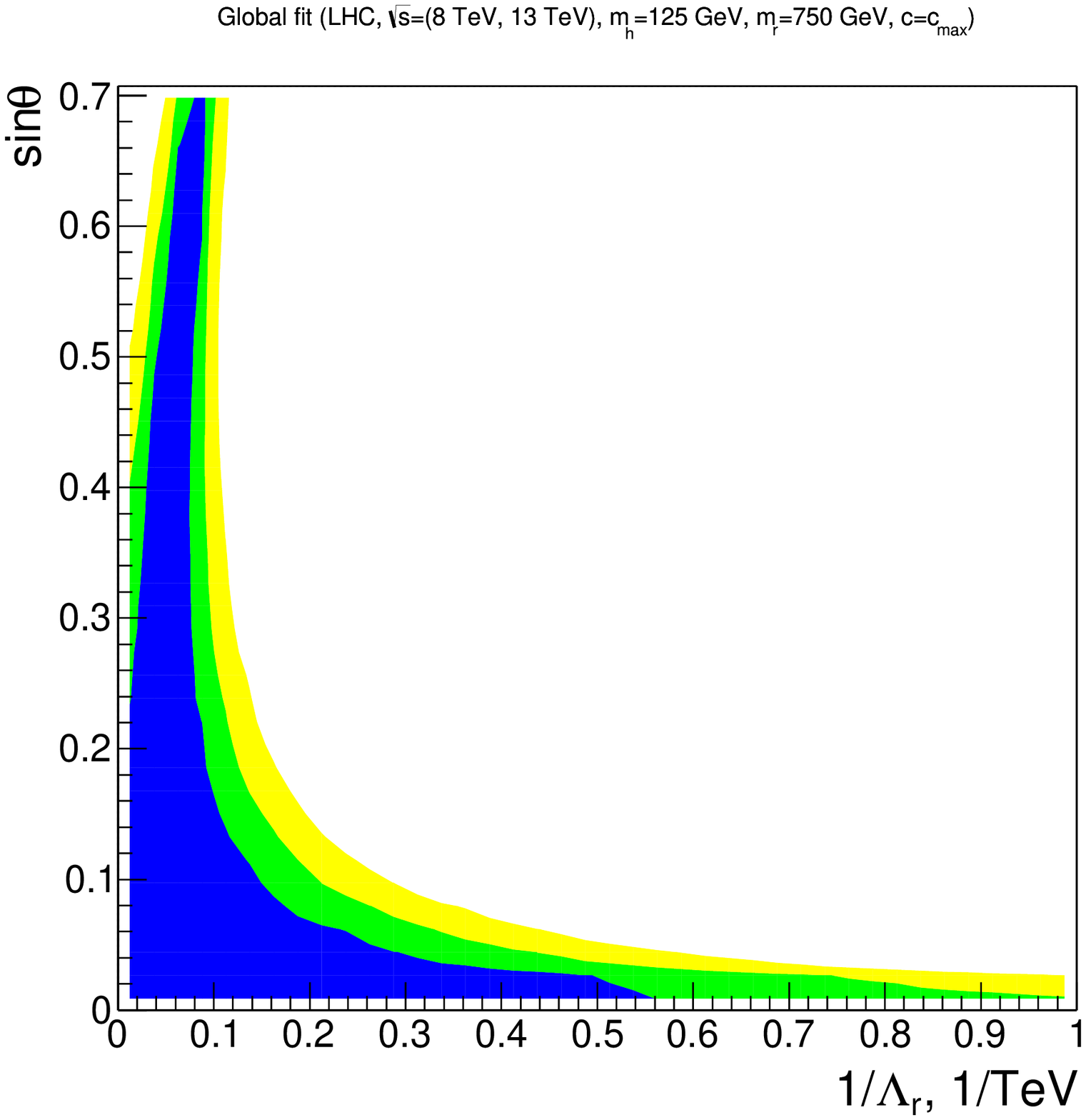}
\caption{ \label{pic8} \footnotesize  Exclusion contours for the
global $\chi^2$ fit in the $(\sin\theta, 1/\Lambda_r)$ plane for
the LHC at $\sqrt{s}=$7, 8 and 13 TeV and  $m_h$=125 GeV,
$m_r$=750 GeV, $c=c_{max}$. The dark, medium and light shaded
areas correspond to CL of the fit 65\%, 90\% and 99\%
respectively.}
\end{minipage}

\end{center}
\end{figure*}
One should note that the parameter region allowed by the signal
strength measurements for 125 GeV boson at 8 TeV LHC energy and
presented in  Fig.~\ref{pic7} includes the above mentioned region,
where the two-photon cross section has a maximum. The inclusion of
the new measurements at 13 TeV for the 125 GeV boson and the 750
GeV possible resonance into the overall fit  leads  to a minor
modification of the allowed region of the model parameter space as
demonstrated in  Fig.~\ref{pic8}.

However, the interpretation of the observed excess as the
radion-dominated state is very problematic or even impossible in
the simplest variant of the discussed brane-world models, where
only the gravitational degrees of freedom are allowed to propagate
in the bulk. Indeed, as one can see in  Fig.~\ref{pic6}, the cross
section has a maximum of about 0.14 fb, which is by a factor of
$50 \div 100$ smaller than what is needed to achieve the observed
level of the cross section for

\begin{figure*}[!h!]
\begin{center}

\begin{minipage}[t]{.45\linewidth}
\centering
\includegraphics[width=80mm,height=70mm]{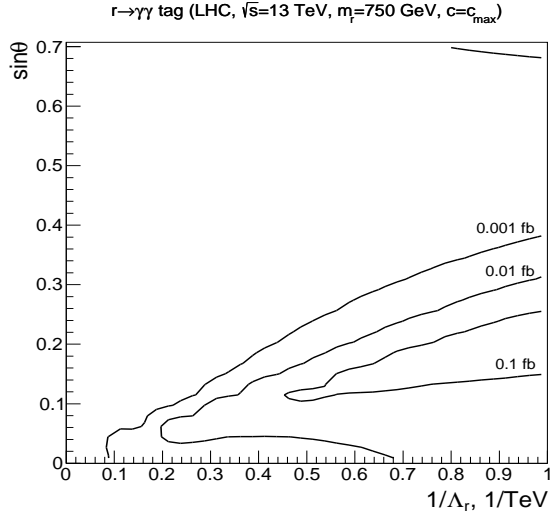}
\vspace{-0.5cm}
\caption{ \label{pic9} \footnotesize Equal value
contours for the production cross section of the radion-dominated
state with mass 750 GeV  as functions of the parameters $\sin
\theta$ and $1/\Lambda_r$.}
\end{minipage}

\end{center}
\end{figure*}
\noindent the 750 GeV excess. In other areas of the  parameter
space the production cross section gets even smaller as one can
see in Fig.~\ref{pic9}, where equal value contours for the cross
section are shown.

Thus, one has to conclude that in the simplest variant of the
model under consideration, where only the  gravitational and
stabilizing scalar fields propagate in the bulk, the observed
excess cannot be understood as a radion-dominated resonance. This
is in accord with the results of paper \cite{Toharia:2008tm},
where the production and two-photon decay of a heavy
radion-dominated state was considered in the unstabilized RS model
with the Higgs-curvature term  and the SM fields on the brane. As
we have mentioned in Section 2, our effective Lagrangian
(\ref{h-r-Lagrangian}) for small values of the parameter $c$ gives
the same coupling of the radion-dominated state to  the  conformal
anomaly of massless vector fields as in paper
\cite{Toharia:2008tm}, which turns out to be dominating for the
small values of the mixing angle, and the corresponding contact
diagrams give the main contribution to the amplitude.

 A similar conclusion about the impossibility to explain the 750
GeV excess by a mixed Higgs-radion state was reached in papers
\cite{Gupta:2015zzs,Megias:2015ory}. In the present paper, in
addition to these previous studies, we analyzed the influence of
the tower of the higher KK scalar states having  performed the
statistical analysis of the allowed parameter space and having
shown that the inclusion of these  KK  states does not change the
main negative conclusion. As was explained above, the obtained
effective Lagrangian of the model  is rather general and therefore
the conclusion remains valid for several other extensions of the
SM by a singlet real scalar field.

In order to increase the two-photon signal rate some other heavy
particles should propagate in the gluon-gluon-scalar and
photon-photon-scalar loop vertices, the scalar in our case being
the Higgs-dominated  or the radion-dominated states. Such heavy
particles in brane-world models could be the excited KK states of
those SM fields, which are also allowed to propagate in the
multidimensional bulk.   The contribution of the bulk field
excitations  to the production cross section of a heavy
radion-dominated state with mass up to 350 GeV   was  discussed in
paper \cite{Toharia:2008tm}. It was  shown there that it can
increase the cross section up to 10 fb, although it is not quite
clear, whether this result can be extrapolated to  the 750 GeV
radion-dominated state. In fact, in paper \cite{Gunion} it is
claimed that the bulk field excitations in the Randall-Sundrum
background can really give the necessary enhancement of the
production  cross section of this state. However, the calculation
of the excited state contributions in stabilized brane-world
models is not straightforward, in particular because of the
excited state impact on the energy momentum tensor, and needs a
special thorough investigation.

\section{Conclusions}
 In the present paper we have studied the possibility of
interpreting the 750 GeV diphoton  excess that has been recently
presented by the LHC as the production of a radion-dominated state
in a stabilized brane-world model. The Higgs-dominated state is
fixed to be the 125 GeV boson. For our investigation we used a
very general effective Lagrangian that describes the low-energy
interactions of the scalar mixed states in the Standard model
extended by a scalar singlet including  the remaining
contributions from the integrated  out possible tower of heavier
scalar states. Our calculations show that although one can rather
well fit the signal strengths in a certain region of the model
parameter space, the resulting two-photon production cross section
is  $50 \div 100$ smaller than what is needed to achieve the
observed level of the cross section. This means that the
stabilized brane-world models with only gravity and the
Goldberger-Wise field propagating in the bulk, as well as the
other models described by  special cases of the considered
effective Lagrangian, are unable to explain the excess.

\section{Acknowledgements}
The authors are grateful to M. Smolyakov for useful discussions.
The work was supported by  grant 14-12-00363 of the Russian
Science Foundation.

\end{document}